\documentclass{amsart}

\usepackage{amssymb}
\usepackage{graphicx}

\newcommand*{\jb}{\mathbf{j}}
\newcommand*{\xb}{\mathbf{x}}
\newcommand*{\nb}{\mathbf{n}}
\newcommand*{\ab}{\mathbf{a}}
\newcommand*{\eb}{\mathbf{e}}
\newcommand*{\omegab}{{\boldsymbol{\omega}}}
\newcommand*{\dd}{\mathrm{d}}

\title{Gravitational waves and the Sagnac effect}

\author{J. Frauendiener}
\curraddr{Department of Mathematics and Statistics, University of Otago, New Zealand}
\address{Institut de Mathématiques, Université de Bourgogne, France}
\email{joergf@maths.otago.ac.nz}
\thanks{This paper is dedicated to the memory of my late father Werner Frauendiener.}

\begin{document}


\begin{abstract}
Light propagating in opposite directions around the same loop in general shows a relative phase shift when recombined. This phenomenon is known as the Sagnac effect after Georges Sagnac who, in 1913, demonstrated with an interferometer on a rotating table that the phase shift depended on the angular velocity of the table. In previous work we have given a very general formula for the Sagnac effect, valid in full general relativity. The relativistic effect not only contains the `classical' contribution from the rotation of the laboratory but also contributions due its acceleration and due to incoming gravitational waves. Here, we point out a major  consequence of this gravitational effect which may have implications for third generation gravitational wave detectors. We describe an ``antenna'' design which picks out specific components of the Weyl tensor describing the incident gravitational waves.
\end{abstract}
\maketitle

\section*{Introduction}

In general relativity, a laboratory is modeled as a time-like world-line to measure the proper time $t$ passing in the lab together with a set of space-like mutually orthogonal vectors $(\eb_1,\eb_2,\eb_3)$ attached at each point of the world-line~\cite{MTW}. This reference frame indicates the orientation of the lab in space-time at each instant of time. The lab may rotate, it may be accelerated and it may travel through an arbitrarily curved region of space-time. Introducing (generalized) Fermi coordinates $(t,x^1,x^2,x^3)$ adapted to the lab~\cite{Manasse:1963ha,Synge:1976vf} one can describe the geometry of space-time by means of its metric $g_{ab}$ with respect to these coordinates by

\[
  \begin{aligned}
      g_{00} &= 1 - 2 a_l x^l + 3 (a_mx^m)^2 + \omega_{im} \omega^i{}_n x^m x^n + R_{m0n0} x^m x^n + O(x^3),\\
    g_{0k} &= \omega_{kl} x^l  + \frac23  R_{m0nk} x^m x^n+ O(x^3),\\
    g_{kl} &= -\delta_{kl} + \frac13 R_{mlnk} x^m x^n+ O(x^3).
\end{aligned}
\]
Here, $a^i$, $\omega^i{}_k$ and $R_{m0n0}$ etc. are components of the acceleration and the angular velocity of the lab and of the Riemann tensor of the space-time. These quantities depend on $t$ and the expressions are valid up to the given order in the spatial coordinates.

Suppose two photons travel around a closed loop $C$ which has no self-intersections. It can be given in parametrised form as $x^i(s)$. The photons start at the same time at the point $Q=x^i(0)$, returning back to $Q$ at different times depending on the travel direction. In~\cite{Frauendiener:2018gx} we derived the general formula for the difference in the arrival times of the photons. This formula is not immediately useful since it involves the solution of a differential equation along the path. However, with the very reasonable assumption that the travel time of the photons is negligibly small compared to the time scales of changes in the lab motion and the surrounding curvature one can derive the succinct formula 

\[
  \Delta T = -2 \int_C \frac{g_{0i}}{g_{00}}\,\dd x^i.
\]
Using the Stokes theorem we can recast this line integral as a surface integral over a surface~$S$ which is bounded by the curve $C$. Inserting the expression for the metric in terms of the Fermi coordinates one obtains three terms contributing to the time difference. To discuss them we use the usual 3-vector notation $\ab$ and $\omegab$ for the acceleration and the angular velocity and we use the position vector $\xb$ and the vector $\nb$ normal to the surface $S$. Then, the first term becomes

\[
  \Delta_\omega T = 4 \int_S \omegab \cdot \nb\, \dd^2 S.
\]
This is the classical Sagnac effect as first described by Sagnac~\cite{Sagnac:1913tx,Sagnac:1913tz} expressed as the `rotation flux' through the surface $S$. It is proportional to the magnitude of the angular velocity but it also depends on its direction in relation to the surface $S$ and therefore to the curve $C$. In fact, by considering different shapes of $C$ one can construct different `antennas', i.e., configurations with different directional dependence. For instance, the curve which is described by the seam of a tennis ball is insensitive to rotations around the two axes piercing the opposite lobes but can detect rotations around the third axis. This contribution is translation invariant.

The second term depends on the rotation as well as the acceleration:

\[
    \Delta_a T = 4  \int_S (\ab \cdot \omegab) (\nb\cdot \xb) - 3  (\ab \cdot \xb) (\omegab\cdot\nb)\,\dd^2S.
\]
The appearance of $\xb$ shows that this contribution is not translation invariant, the time difference also depends on the position of the loop. This term can, at least in principle, be used to detect the acceleration of the lab in relation to the rotation axis and the orientation of the curve.

The third term is due to the curvature of the space-time and has two separate parts. One of the pieces is caused by the Weyl tensor and, hence, is related to gravitational waves while the other part comes from the Ricci tensor and, therefore due to the Einstein equation, is caused by the matter content of the space-time:

\[
  \Delta_RT = 4 \int_S \nb \cdot \mathbb{B} \cdot \xb\, \dd^2S - \frac{16\pi G}{c^4} \int_S \nb \cdot (\xb\times \jb)\, \dd^2S.
\]
In the first term, $\mathbb{B} =(B_{ij})$ is a symmetric, trace-free $3\times3$-matrix which describes the magnetic part of the Weyl tensor. It contains the information of gravitational waves propagating in the three spatial directions together with their two independent polarisation states. In the second part we find the momentum density $\jb$ of the matter so that this term is caused by the flux of the material angular momentum density through $S$.

Both parts of the gravitational contribution depend on the position of the curve. For the remainder we focus on the gravitational wave part. To give a very simple example we consider a closed path without self-intersections in a plane with normal vector $\nb$ through a point $\xb_0$. We choose the axes so that $\nb = \eb_3$, i.e., it points in the positive $z$-direction. Then the points on the plane can be written in the form

\[
  \xb = \xb_0 + u \eb_1 + v \eb_2.
\]
and the curve itself is specified by a parametrization $(u(t),v(t))$. The time difference for photons traveling along that path is then obtained from the integral

\[
  \Delta_RT = 4 \int_S \nb \cdot \mathbb{B} \cdot \xb\, \dd^2S = 4 A (\nb\cdot \mathbb{B}\cdot \xb_0  + \nb\cdot \mathbb{B}\cdot \eb_1 u_0  + \nb\cdot \mathbb{B}\cdot \eb_2 v_0 )
\]
where $A$ is the (signed) area enclosed by the curve. Here, we have defined $u_0=A^{-1}\int_S u \dd u \dd v$ and similarly for $v_0$. These define the `center of mass' of the area surrounded by the curve. Choosing this point as $\xb_0$ we can obtain the simple formula 

\[
  \Delta_RT = 4 A (\nb\cdot \mathbb{B}\cdot \xb_0 )
\]
for the time difference along a simple path in a plane with normal vector $\nb$ passing through the point $\xb_0$ which is the center of mass of the area $A$ enclosed by the curve.

\begin{figure}[h]
  \centering
  \includegraphics[width=0.3\linewidth]{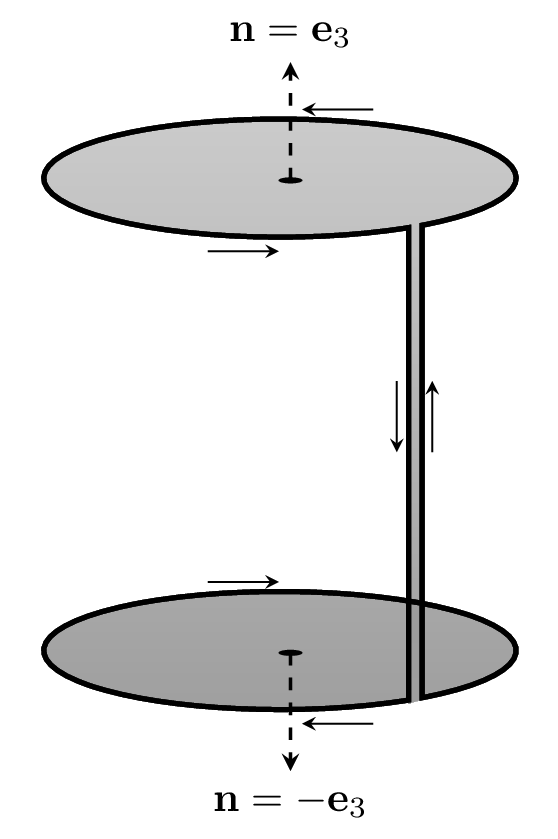}
  \caption{A simple configuration of loops for a Sagnac detector for gravitational waves}
  \label{fig:loops33}
\end{figure}

For this simple system the time difference depends on the location $\xb_0$ of the loop. However, we can combine such loops and obtain more complicated configurations with better behaviour. One example is shown in Fig.~\ref{fig:loops33}. Other possibilities exist.
The configuration consists of two identical loops which are almost closed and connected in such a way that a photon which travels counterclockwise in the upper loop around the $z$-axis will loop around in the clockwise direction in the lower loop. In the ideal situation, the vertical strip can be made arbitrarily small compared to the area enclosed by each loop. The loops are positioned so that the upper loop is centred around $\xb_0 = (a,b,c)$ and the lower one is centred at $\xb_1 = (a,b,-c)$. The orientation of the loops has the consequence that the normal vector is oriented along the positive $z$-axis in the upper part and points in the opposite direction in the lower part of the path.

The net effect of the time difference between two counter propagating photons can then be obtained by simply adding the contributions from each loop taking into account their different orientations. The result is
\[
  \Delta T = 4 A (\nb\cdot \mathbb{B}\cdot \xb_0 + \nb\cdot \mathbb{B}\cdot \xb_1) = 4 A B_{33} h.
\]
Here, we introduced the height $h=2c$ of the configuration and we denote by $A$ the area of one loop. This ``antenna'' is sensitive to exactly one component of the magnetic part of the Weyl tensor. As a bonus, we find that it is not sensitive at all to the classical Sagnac effect when the vertical area is made zero.

\begin{figure}[h]
  \centering
  \includegraphics[width=0.5\linewidth]{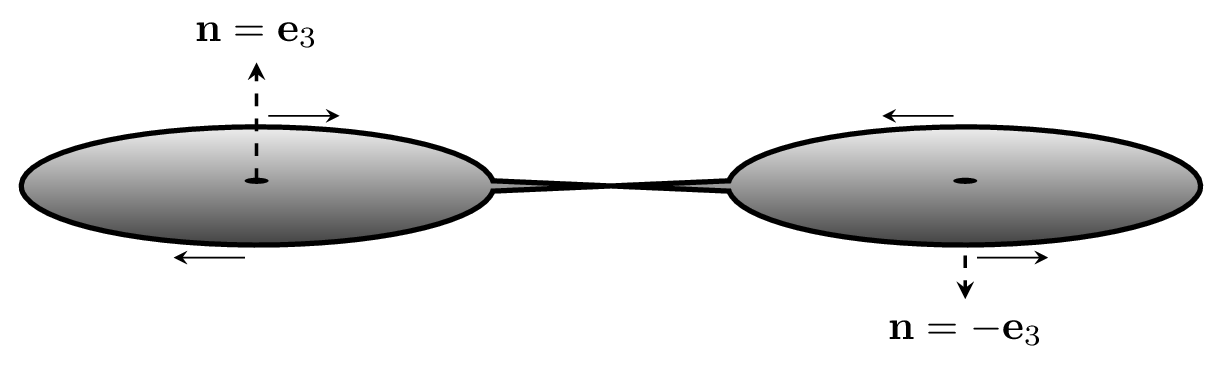}
  \caption{A loop configuration to detect an off-diagonal element of the magnetic part of the Weyl tensor}
  \label{fig:loops13}
\end{figure}

In a similar way we can construct a configuration which is sensitive to an off-diagonal element of the Weyl tensor, see Fig.~\ref{fig:loops13}.
Here, we have assumed the loops to lie in the same plane. They are traversed with different orientations so that $\nb=\eb_3$ for one loop and $\nb=-\eb_3$ for the other. The loops are centered around the points $\xb_0=(-a,0,0)$ and $\xb_1=(a,0,0)$, respectively,  and connected along two crossing lines. For this system we obtain for the time difference between two counter propagating photons the value
\[
  \Delta T = 4 A (\nb\cdot \mathbb{B}\cdot \xb_0 + \nb\cdot \mathbb{B}\cdot \xb_1) = 4 A B_{13} L
\]
where we have introduced $L=2a$, the distance between the centres of the two loops. This shows that we can --- at least in principle --- devise configurations which are able to pick up all the components of the Weyl tensor. In particular, a combination of such loops can be arranged in such a way that the antenna is also sensitive to the polarisation of the incident gravitational wave.

As a final remark, we point out that this way of detecting gravitational waves is, in some sense, dual to the detectors in current use which are based on a Michelson type interferometer. These detect the wave form due to the geodesic deviation equation which is driven by the electric part of the Weyl tensor. This, and the particular design of Sagnac antennas may be of interest to the current discussion of third generation gravitational wave detectors, see~\cite{Bond:2017vy}.


\section*{Acknowledgments}
The author is grateful to Prof. Ezra Newman and Prof. Rainer Weiss for comments.

\providecommand{\newblock}{}


\end{document}